\documentclass[prl,aps,amssymb,showpacs,twocolumn]{revtex4}

\usepackage{amsmath}
\usepackage{amssymb}
\usepackage{amsthm}
\usepackage{amsfonts}
\usepackage{algorithmic}
\usepackage{enumerate}
\usepackage{latexsym}
\usepackage[dvips]{graphicx}

\newcommand{\beq}{\begin{equation}}
\newcommand{\eneq}{\end{equation}}
\newcommand{\bed}{\begin{displaymath}}
\newcommand{\ened}{\end{displaymath}}
\input{epsf}

\begin{document}

\tolerance 10000


\title{Roton Induced Modulations in Underdoped Cuprates as a Signature
of Incipient Electronic Order}

\author { Zaira Nazario$^\dagger$ and 
David I. Santiago$^{\dagger, \star}$ }

\affiliation{ $\dagger$ Department of Physics, Stanford University,
             Stanford, California 94305 \\ $\star$ Gravity Probe B Relativity 
             Mission, Stanford, California 94305}
\begin{abstract}

\begin{center}

\parbox{14cm}{ In the last years there have been measurements of
energy independent modulations in underdoped cuprates via scanning
tunelling microscopy. These modulations of around 4 lattice constants
occur near vortex cores in samples with and without superconductivity,
and in the pseudogap regime of some superconducting samples. Recently,
it was proposed that the origin of the modulations is electronic charge ordered
phases intrinsic
to doped Mott insulators. We show that even if the system has 
not charge ordered, proximity to 
such an order will induce  modulations at the same ordering wavevector. Such
modulations are a signature of the real or incipient order. We construct
ground state wavefunctions that include strong fluctuations of the
competing order without having ordered. They exhibit deep roton
minima at wavevectors of the reciprocal lattice of the incipient order
due to proximity to the phase transition into the charge ordered
state. Such minima, in turn, produce strong modulations. Moreover, the effects 
are generic and are present whenever an
electronic system is near a crystalline phase. This leads to
experimental signatures even if the electronic system has not solidified. The
signatures will be ubiquitous near the dopings that stabilize the 
electronic order.}

\end{center}
\end{abstract}


\maketitle

In the present note we would like to point out that the recent
proposed explanation\cite{dhl} for STM modulations at certain doping
dependent wavevectors in samples with and without superconductivity,
and around vortex cores\cite{exp1} is of a quite general nature. The 
modulations around the vortex core were predicted in the context of the $SO(5)$
picture of high T$_c$\cite{sc}, but their nature is more general than the 
specific model. We
show more directly how the physics comes about by means of a quantum mechanical
density functional formalism. The proposal\cite{dhl} is of an enhanced
charge susceptibility due to charge order or to a deep roton minimum
at certain wavevectors. We show that roton induced modulations are always 
present near a solidification quantum transition regardless of whether the 
electron system has charge ordered or not. Therefore the experimental 
signatures should be quite generic and rather common as long as one is
close to an electronic crystal phase.

That roton physics follows on quite general grounds can be inferred
from the experimental properties of bosonic systems. Helium is a
superfluid at zero temperature, but it is an almost solid barely
melted by quantum fluctuations. An indication of its proximity to the
solid phase is the deep roton minimum in the excitation
spectrum\cite{landau,feyn1} or, equivalently, the peak in the
structure factor at the same wavevector. On the other hand,
artificially engineered superfluids such as atomic BECs\cite{bec1},
which are not near a solid transition by experimental design, do not
have a peak in the structure factor\cite{kett} and hence they will not
have a roton excitation unless they are tuned near a Mott transition.

In the case of the cuprates there is a plethora of possible Coulombic
stabilized insulating phases: the low doping antiferromagnetic
insulator, stripes\cite{str}, hole Wigner crystals\cite{dhl} and even
a semimetallic orbital antiferromagnet\cite{ddw}, among
others. Whenever one of these phases {\it with charge order} (stripes,
Wigner crystals, charge density waves) becomes competitive in energy
with the superconductivity, an incipient Bragg peak will start
forming, producing enhanced fluctuations at appropriate wavevectors
and thus a spike in the structure factor. This spike leads to a roton
minimum in the excitation spectrum \cite{feyn1,girvin}. In fact, the
magnetic neutron $(\pi, \pi)$ mode\cite{kaimer} is such a roton
minimum in the spin rather than charge density excitation spectrum. It
is due to proximity to an electron crystal phase, the
antiferromagnetic insulator, at very low dopings\cite{ddw}.

The Hamiltonian
\beq \label{haml}
\mathcal{H} = \int \left(m \frac{\vec v \cdot \rho \vec v}{2} \right)
d^3r + U[\rho]
\eneq
\noindent describes quantum many particle systems \cite{landau}. $\vec
v \cdot \rho \vec v / 2$ is the kinetic energy operator, $U[\rho]$ is
the potential energy operator, which in general can be a functional of
the density operator. The density operator in first quantized notation
is 
\beq
\rho(\vec r) = \sum_i \delta(\vec r - \vec r_i)
\eneq

\noindent with $\vec r_i$ being the position of the ith particle. The
velocity operator is in first quantized notation 
\beq
\vec v(\vec r) = \sum_i \left[ \frac{\vec p}{2m}\delta(\vec r - \vec
r_i) + \delta(\vec r - \vec r_i)\frac{\vec p}{2m}\right]
\eneq
\noindent with commutation relation 
\beq \label{com}
\left[ \rho(\vec r\;')\;, \vec v(\vec r) \right] = i \frac{\hbar}{m}
\nabla \delta(\vec r\;' - \vec r)
\eneq
\noindent The ground state of the Hamiltonian (\ref{haml}) could be a
quantum liquid or solid depending on the interaction. Since our main
concern is how a crystallization transition is approached, we will
suppose it to be a liquid, of course a superconductor for the case of
the cuprates, but our considerations are more general.

We study the nature of the density excitation spectrum of the fluid on quite
general grounds by the methods of Landau and the Russian school\cite{abri}. We 
suppose the density to have a well defined average
which we take to be uniform for simplicity. This will not change the
nature of the considerations, although in real life the density can be
modulated by the lattice. The average velocity is zero as we consider
a system at rest. We Fourier expand the density and velocity operators
about their averages:
\begin{align}
\rho(\vec r) = \rho_0 + \frac{1}{N} \sum_{\vec k} \rho_{\vec k}
e^{i\vec k \cdot \vec r} \\
\vec v(\vec r) = \frac{1}{N} \sum_{\vec k} \vec v_{\vec k} e^{i\vec k \cdot
\vec r}
\end{align}
\noindent where we are using the lattice normalization, the number of
sites $N$ instead of the volume $V$. Quantum mechanically, the
velocity is proportional to the gradient of the local phase variable
of the particle, $\vec v(\vec r)=\hbar/m\nabla\theta$, or in Fourier
components
\beq
\vec v_{\vec k}=\frac{i\hbar\vec k}{m} \theta_{\vec k}
\eneq
\noindent The density velocity commutation relation (\ref{com}) implies
the well known density phase commutation relation
\beq
[\; \rho_{\vec k}\; , \;  \theta_{-\vec k'} \; ] = i \delta_{\vec k \vec k'}
\eneq
\noindent i.e. $\hbar\theta_{-\vec k}$ is the momentum conjugate to
the density fluctuation $\rho_{\vec k}$. The Hamiltonian (\ref{haml})
thus becomes 
\beq \label{ham2} 
\mathcal H = U(\rho_0) + \frac{1}{N} \sum_{\vec k} \left\{
\frac{\rho_0\hbar^2 k^2}{2m}|\theta_{-\vec k}|^2 + \frac{1}{2}
\left(U_{\vec k} + \frac{\hbar^2 k^2}{4m\rho_0}\right) |\rho_{\vec
k}|^2 \right\}
\eneq
\noindent where $U_{\vec k}$ is the Fourier transform of the
interaction and $(\hbar^2 k^2)/(4m\rho_0)|\rho_{\vec k}|^2$ is the
elastic energy associated with changing the density. This
last term comes from acting the Hamiltonian on the wavefunction and
making sure that one keeps careful track of both density and phase
degrees of freedom\cite{dhl2} in the full quantum mechanical kinetic
energy term.

We see that the quantum liquid is a collection of harmonic oscillators
in momentum space for the density fluctuations. The mass of the
oscillators is given by 
\beq
M_{\vec k} \equiv \frac{m}{\rho_0 k^2}
\eneq
\noindent and spring constant by
\beq
K_{\vec k} \equiv U_{\vec k} + \frac{\hbar^2 k^2}{4m\rho_0} \; .
\eneq 
\noindent The density energy excitation spectrum of the quantum liquid
is
\beq
E_{\vec k} = \hbar \omega_{\vec k} (n + 1/2) 
\eneq
\beq 
\omega_{\vec k}^2 = \frac{K_{\vec k}}{M_{\vec k}} = \frac{\rho_0
k^2}{m} \left(U_{\vec k} + \frac{\hbar^2 k^2}{4m\rho_0} \right)
\eneq
\noindent The ground state energy is given by $U(\rho_0) + \sum_{\vec k} \hbar
\omega_{\vec k} / 2$. We note that the Virial theorem implies that
\beq \nonumber
\frac{1}{2} \hbar \omega_{\vec k} N = \left( U_{\vec k} +
\frac{\hbar^2 k^2}{4m\rho_0} \right) \langle |\rho_{\vec k}|^2 \rangle
\eneq
\beq \label{ex}
\hbar \omega_{\vec k} = \frac{\hbar^2 k^2}{2 m S(\vec k)}
\eneq
\noindent where the structure factor is defined by 
\beq 
S(\vec k) =
\frac{\langle |\rho_{\vec k}|^2 \rangle}{N \rho_0} 
\eneq
\noindent which is, of course, the Fourier transform of the
density-density correlation function. For the quantum liquid we thus have
\beq
S(\vec k) =\frac{\hbar k}{2m}\frac{1}{\sqrt{\frac{\rho_0}{m} \left(U_{\vec k}+
\frac{\hbar^2 k^2}{4m\rho_0}\right)}}
\eneq

\noindent The ground state energy will have the density oscillators unexcited.
Hence, if the ground state wavefunction of the electronic system 
{\it without density correlations}  is $|\psi_{GS}\rangle$, the ground state 
wavefunction {\it with density correlations} is
\beq \label{gs}
|\psi\rangle = \exp\left \{ -\frac{1}{2}\sum_{\vec k} 
\sqrt{\frac{m}{\rho_0 \hbar^2 k^2 }
\left(U_{\vec k} + \frac{\hbar^2 k^2}{4 m \rho_0}\right)} |\rho_{\vec
k}|^2 \right\}|\psi_{GS} \rangle \; .
\eneq
\noindent The factor in front appears because it is the ground state
wavefunction of harmonic oscillators. The ground state wavefunction
has the oscillators shaking as little as possible. We see that the
factor suppresses density fluctuations. The wavefunction is
essentially exact to quadratic order in the density fluctuations. This
way of including correlations was pioneered by Bohm and
Pines\cite{rpa} and is known colloquially as the RPA. This correlated
ground state wavefunction is a Jastrow type wavefunction analogous to
the Laughlin Fractional Quantum Hall wavefunction\cite{dhl2}.

One could play the same game as for the ground state wavefunction (\ref{gs})
and write down the excited state wavefunctions for the density oscillators.
On the other hand, these collective density oscillations are mostly not good
approximations to eigenstates of the system since they
will readily decay into particle-hole pairs except at the longest wavelengths.
Such long wavelength oscillations are plasmons, for which the present 
consideration leads readily  to $\omega_{\vec
k}^2= 4 \pi e^2 \rho_0/m$, as for the electron Coulomb interactions we
have $U_{\vec k} = 4\pi e^2 / k^2$ which dominate as $\vec k \rightarrow 0$.

The ground state wavefunction (\ref{gs}) does not contain any electronic 
crystalline order as $\langle \rho_{\vec k} \rangle = 0$ for all $\vec k$.
When the electron system solidifies, either spontaneously or lattice induced, 
it will exhibit Bragg peaks at the reciprocal lattice vectors corresponding to 
the electron crystal. These peaks mean that $\langle \rho_{\vec Q} \rangle \neq
0$. We are interested in the case when the electron system solidifies 
spontaneously, and in particular, in its behavior as it approaches such a 
point. 
\begin{figure}[ht]
\centering
\rotatebox{270}{
  \resizebox{4.8cm}{!}{%
    \includegraphics*{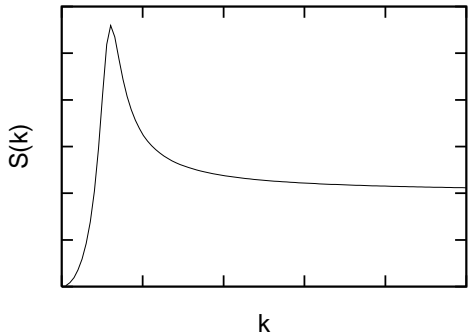}}}
\caption{Structure factor vs. momentum. The roton maximum is easily
visible.}
\end{figure}

As the electron fluid approaches spontaneous crystallization with
reciprocal wavevector $\vec Q$ for the incipient order, there will be
enhanced density fluctuations $\langle |\rho_{\vec Q}^2| \rangle$ at
such a wavevector. This follows because proximity to the
crystallization softens the density oscillators of the electron
liquid, $\vec K_{\vec Q} \rightarrow 0$, which makes $\omega_{\vec Q}
\rightarrow 0$. Hence the interaction $U_{\vec k}$ has a minimum at
$\vec Q$. The softening of the density mode thus produces a peak in
the structure factor, $S(\vec k)$, that diverges as one approaches the
point when the system solidifies. This peak will drag the spectrum
down according to the equation (\ref{ex}) creating a roton
minimum\cite{feyn1,girvin} that can be thought of as an exciton. We
thus have as a simple phenomenological model that captures the physics
\begin{align}
\nonumber U_{\vec k} &= \frac{4 \pi e^2}{k^2} + \alpha
\frac{\hbar^2}{4 m \rho_0} (\vec k - \vec Q)^2 - \frac{4 \pi e^2}{Q^2} \\
&+ \alpha \frac{\hbar^2 Q^2}{4 m \rho_0} - \alpha \frac{\hbar^2
k^2}{4 m \rho_0} + \Delta_{\text{roton}}
\end{align}

\noindent where nonzero $\alpha$ makes a roton minimum, and
$\Delta_{\text{roton}}$ is the roton gap, which will collapse at the
crystallization transition as the smallness of the gap of such a roton
is a measure of proximity to crystallization. From the form of the
interaction we get the structure factor and the density excitation
spectrum, which we plot in Figures 1 and 2, where the roton wavevector
was chosen arbitrarily as 1, $\alpha$ was chosen as $1/2$, the roton
gap $\Delta_{\text{roton}}$ as $1/4$ and all other dimensionful
constants were chosen to have unit scale.

We consider the case when an external potential $- V_{\vec k}$ acts on the 
electrons thus adding a term
\beq \label{hamext}
\mathcal H_{\text{ext}} = - \sum_{\vec k} V_{\vec k} \rho_{\vec k} \; .
\eneq
\noindent The external potential might be the ionic lattice or it might be
the impurity potential. The effect of the extra term in the ground state
wavefunction is to move the equilibrium position of the density oscillators
from $\langle \rho_{\vec k} \rangle = 0$ to
\beq \label{rho}
\langle \rho_{\vec k} \rangle = \frac{V_{\vec k}}{U_{\vec k} + 
\frac{\hbar^2 k^2}{4m\rho_0}} \; .
\eneq
\noindent We point out that when $V_{\vec k}$ is the lattice potential, this
is just the electron fluid acquiring the periodicity of the underlying lattice
with appropriate Bragg peaks. 
\begin{figure}[ht]
\centering
\rotatebox{270}{
  \resizebox{4.8cm}{!}{%
    \includegraphics*{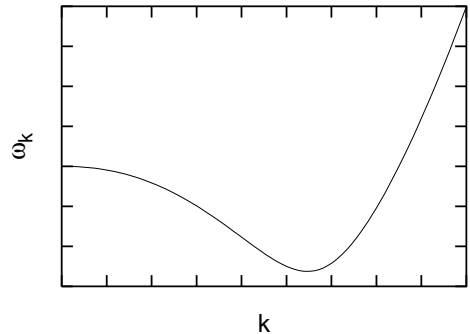}}}
\caption{Density oscillation frequency vs. momentum. The plasmon gap
and the roton minimum are easily visible.}
\end{figure}

When the electron fluid is near a solidification transition with
reciprocal lattice vector $\vec Q$ for the incipient order, there will
be a large enhancement of the density component at or near wavevector
$\vec Q$ due to the roton minimum unless $V_{\vec k} \equiv 0$ for the
region around the minimum. In figure 3 we plot the enhancement of the
density Fourier component for a constant potential which could
correspond to a localized impurity. The scale of the potential in the
figure was chosen arbitrarily as 1. On the other hand, we stress that
the effect is general and the enhancement will be there for the
lattice potential. Quasiparticles will scatter strongly of the roton
enhanced density Fourier components at and around $\vec Q$. These
corroborates the previous suggestions\cite{dhl} and we thus identify
this enhancement with the STM\cite{exp1} bias independent signatures.

We point out that our result shows the generality of the roton induced 
modulations near an electronic crystal phase effect, and that charge order
is not necessary for the observation of the experimental signatures. The 
physics follows because it is very easy to make density fluctuations at the 
roton wavevectors. Since the electron density is very high at roton 
wavevectors, there will be an enhanced charge susceptibility at such
wavevectors leading to modulations\cite{dhl} {\it as long} as the system is
near enough to an electronic crystal phase. The phenomenon  can be observed in 
superconducting samples which might not have the
charge order. For example, in underdoped samples if superconductivity
is lost at a temperature not far above the roton gap, these
modulations will be present. Since $T_c$ collapses with underdoping,
the modulations will be easier to observe for small doping.
\begin{figure}[ht]
\centering
\rotatebox{270}{
  \resizebox{4.8cm}{!}{%
    \includegraphics*{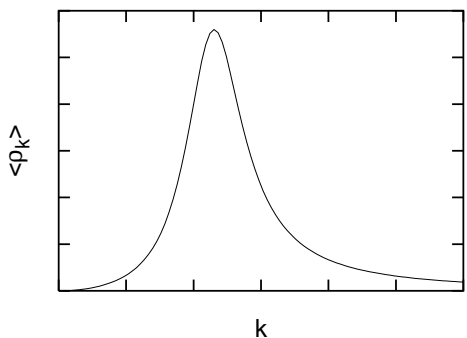}}}
\caption{Density Fourier component vs. wavevector. The enhancements of
the density Fourier components at and around the roton wavevector are
easily visible.}
\end{figure}

Measurements of metallic like conductivity below the Neel
temperature\cite{ando1} in underdoped cuprates, Hall
effect\cite{ando1} transport measurements, and the different
photoemission properties of nodal vs. antinodal
quasiparticles\cite{zx} have as the simplest explanation that there
are two types of electronic carriers in underdoped cuprates. It is
certainly suggestive that the experimental wavevectors for the
modulations are in the antinodal directions\cite{exp1}. The roton model
will be a two fluid system as there will be low energy quasiparticles
near the roton minimum which will behave more like excitations of the
competing insulating order, while low energy quasiparticles coming
from other wavevectors will tend to behave like superconducting
Bogolyubov-BCS excitations.

We finally point out that there are reports of a Boomerang effect in the 
cuprates\cite{boom}. This consists in $T_c$ becoming proportional to 
superfluid density,
$n_s$, in the overdoped regime as a consequence of a decreasing number of
superconducting electrons. This suggests a loss of spectral weight of the 
superconducting phase to a competing order. If the roton minima are indeed
signatures of incipient electronic crystal phases and the competing phase on 
the overdoped side is an electronic crystal, there will be similar 
modulations in the overdoped regime observable in STM experiments.

We would like to mention that recently there was also a proposal that
the STM modulations are a Copper pair crystal\cite{sc2}. On similar
physical grounds as expressed in\cite{dhl}, we believe such a crystal
would not occur, but we emphasize that the two proposals are
differentiable experimentally according to at which doping one
observes specific modulations. We also stress that the physics we have
described is more general than either proposal because as long as
there is an incipient solid order, there will be roton minima at its
reciprocal wavevectors and the concomitant modulations.

{\bf Acknowledgements}: We are indebted to Dung-Hai Lee for extremely
useful discussion and suggestions critical to this work. We thank
Shoucheng Zhang for very useful comments. Zaira Nazario is a Ford
Foundation predoctoral fellow.  She was supported by the Ford
Foundation and by the School of Humanities and Science at Stanford
University. David I. Santiago was supported by NASA Grant NAS 8-39225
to Gravity Probe B.


\begin{thebibliography}{99}

\bibitem{dhl} Henry C. Fu {\it et. al.}, arXiV:condmat/0403001 (2004).

\bibitem{exp1} C. Howald {\it et. al.}, e-print arXiV:cond-mat/0208442
  (2002); J. E. Hoffman {\it et. al.}, Science {\bf 295}, 466 (2002);
  M. Vershinin {\it et. al.}, 10.1126/Science.1093384 (Science Express
  Reports), February 12, 2004.

\bibitem{sc} H. D. Chen {\it et. al.}, Phys. Rev. Lett. {\bf 89}, 137004
(2002).

\bibitem{landau} L. D. Landau, Journal of Physics {\bf V}, 71 (1941).

\bibitem{feyn1} R. P. Feynman, Phys. Rev. {\bf 94}, 262 (1954); R. P. Feynman 
and M. Cohen, Phys. Rev. {\bf 102}, 1189 (1956). 

\bibitem{bec1} K. B. Davis {\it et. al.}, Phys. Rev. Lett. {\bf 75}, 3969 
(1995);  C. C. Bradley {\it et. al.}, Phys. Rev. Lett. {\bf 75}, 1687
(1995); M. H. Anderson {\it et. al.}, Science {\bf 269}, 198 (1995). 

\bibitem{kett} W. Ketterle {\it et. al.}, Phys. Rev. Lett. {\bf 83}, 2876 
(1999).

\bibitem{str} D. Poilblanc and T. M. Rice, Phys. Rev. B {\bf 39}, 9749 
(1989);J. Zaanen and O. Gunnarson, Phys. Rev. B {\bf 40}, 7391
(1989); H. J. Schulz, Phys. Rev. Lett. {\bf 64}, 1445 (1990); V. J. Emery and
S. A. Kivelson, Physica C {\bf 209}, 597 (1989); J. M. Tranquada {\it et. al.},
Nature {\bf 375}, 561(1995);

\bibitem{ddw} S. Chakravarty {\it et. al.}, Phys. Rev. B {\bf 63}, 094503 
(2001).

\bibitem{girvin} S. M. Girvin {\it et. al.}, Phys. Rev. B {\bf 33}, 2481 
(1986). 

\bibitem{kaimer} H. F. Fong {\it et. al.}, Phys. Rev. Lett. {\bf 75},
316 (1995).

\bibitem{abri} A. A. Abrikosov {\it et. al.}, ``Methods of Quantum
Field Theory in Statistical Physics'', Prentice Hall, Englewood
Cliffs, NJ (1963).

\bibitem{dhl2} C. Kane {\it et. al.}, Phys. Rev. B {\bf 43}, 3255 (1991).

\bibitem{rpa} D. Bohm and D. Pines, Phys. Rev. {\bf 80}, 903 (1950);
{\it ibid} {\bf 82}, 625 (1951); {\it ibid} {\bf 92}, 609 (1953);
D. Pines, Phys. Rev. {\bf 92}, 626 (1953).

\bibitem{ando1} Y. Ando, arXiV:cond-mat/0206332 (2002); Y. Ando 
{\it et. al.}, J. Low Temp. Phys. {\bf 131}, 793 (2003); Y. Ando {\it et. al.},
 arXiV:cond-mat/0401034 (2004).

\bibitem{zx} X. J. Zhou {\it et. al.}, arXiV:cond-mat/0403181;
  to appear in Phys. Rev. Lett. (2004).

\bibitem{boom} C. Niedermayer {\it et. al.}, Phys. Rev. Lett. {\bf 71}, 1764 
(1993); J. L. Tallon {\it et. al.}, Phys. Rev. Lett. {\bf 74}, 1008 (1995);
C. Bernhard {\it et. al.}, Phys. Rev. B {\bf 52}, 10488 (1995); R. Puzniak
{\it et. al.}, Phys. Rev. B {\bf 53}, 86 (1996).

\bibitem{sc2}  H. D. Chen {\it et. al.}, arXiV:cond-mat/042323 (2004).

\end{thebibliography}
\end{document}